\documentclass[twocolumn,showpacs,superscriptaddress,prl,aps,floatfix]{revtex4}
\usepackage{graphicx}
\usepackage{rotating}
\usepackage{amsmath}

\begin{document}

\title{Triangular Mott-Hubbard Insulator Phases of Sn/Si(111) and 
Sn/Ge(111) Surfaces}

\author{G.~Profeta}

\affiliation{CNISM and Dipartimento di Fisica, Universit\`a degli Studi di
L'Aquila, I-67010 Coppito (L'Aquila) Italy}

\author{E.~Tosatti}
\affiliation{International School for Advanced Studies (SISSA), and INFM
Democritos National Simulation Center, Via Beirut 2-4, I-34014 Trieste,
Italy}
\affiliation{International Centre for Theoretical Physics
(ICTP), P.O.Box 586, I-34014 Trieste, Italy}

\begin{abstract}
The ground state of Sn/Si(111) and Sn/Ge(111) surface $\alpha$-phases is 
reexamined theoretically, based on $ab-initio$ calculations where 
correlations are approximately included through the orbital 
dependence of the Coulomb interaction 
(in the local density + Hubbard U approximation). The effect of correlations
is to destabilize the vertical buckling in Sn/Ge(111) and to make the 
surface magnetic, with a metal-insulator transition for both systems. This signals the
onset of a  stable narrow gap Mott-Hubbard insulating state, in agreement with 
very recent experiments. Antiferromagnetic exchange is proposed to be
responsible for the observed $\Gamma$-point photoemission intensity, as well as for
the partial metallization observed above above 60 K in Sn/Si(111).
 Extrinsic metallization 
of Sn/Si(111) by, $e.g.$ alkali doping, could lead to a novel 2D triangular 
superconducting state of this and similar surfaces.
\end{abstract}
\pacs{71.30.+h, 73.20.-r, 73.20.At, 75.70.Rf, 75.70.Ak, 71.27.+a,74.78.Na}


\maketitle

Metal-insulator transitions and strongly correlated states of electrons in 
the genuinely two dimensional (2D) ``dangling bond'' 
surface states of semiconductors have long been sought after \cite{pwa} but 
seldom realized and characterized. On these surfaces, band physics 
usually takes over, driving large structural reconstructions, wich remove metallicity 
with a strong energy gain \cite{zangwill}. The resulting passivation 
of surface states unfortunately also removes alternative and
more interesting phases, including charge density waves (CDW) and 
spin density waves, 
Mott-Hubbard insulators (MIs), and possibly 2D superconductivity \cite{bertel}.

Plummer and others \cite{carpinelli, carpinelli1, weitering,johansson, petersen} 
however called attention to the $\alpha$-phase surfaces, obtained by 
covering an ideal metallic (111) semiconductor surface with a 
$\sqrt3\times\sqrt3R30^{\circ}$ ($\sqrt3$) triangular 
array of group IV adatoms. These systems, where adatoms lie as far
apart as $\sim$ 7 \AA , possess a very narrow half-filled adatom surface 
state band, making them ideally prone to various instabilities and 
to strong correlations. A low temperature CDW-like reversible  
$3\times3$ periodic surface adatom distortion was indeed reported in metallic 
Pb/Ge(111) and Sn/Ge(111) \cite{carpinelli1, petersen} whereas undistorted triangular
MI states appeared to prevail in isoelectronic surfaces
like Si/SiC(0001) \cite{johansson} and  K:Si(111)$\sqrt3 -B$ \cite{weitering}. 

Density functional calculations in the local density approximation (LDA) supported
this diversity of behavior, indicating that a large $\sim$ 0.3 \AA\ periodic ``up-down'' 
distortion \cite{avila, perez, degironc} akin to a valence disproportionation 
\cite{ballabio} should be the ground state of Sn/Ge(111), against an undistorted magnetic insulator 
prevailing in a large gap system like Si/SiC(0001) 
\cite{northrup, hellberg,santoro}.
Interestingly, the intermediate case of Sn/Si(111) failed to fall neatly on 
either side of this divide. Within LDA, this surface is equally close to 
undistorted magnetism (evolving to an insulator as we shall see later) 
as it is to a $3\times3$ distorted metal \cite{perez}; but neither 
state is actually stable in the strain free surface \cite{ballabio}. 
As we will show presently, $T=0$ LDA seems paradoxically to describe better 
the behavior of these surfaces at {\em non-zero} temperature, whereas 
a better account of correlations is needed to describe their actual ground state. 

Systems realizing in 2D a spontaneous transition between these 
two types of state, namely the distorted (or undistorted) nonmagnetic  
band metal and the magnetic undistorted MI are very interesting to pursue. 
Among other things the latter constitute the building block of many important 
strongly correlated systems, including (in the square lattice version) cuprate superconductors. 
Model studies do indicate that distorted metal--undistorted MI transitions 
are to be expected in the present system as a function of parameters \cite{santoro}. Correlations 
in Sn/Si(111) and Sn/Ge(111) surfaces were discussed by
 Flores {\em et al.} \cite{flores}, who however concluded against a transition 
to a MI ground state. In fact, until 
recently no such transition was actually observed in $\alpha$-phases.

\begin{figure}
\begin{center}
\includegraphics[width=0.5\textwidth,clip]{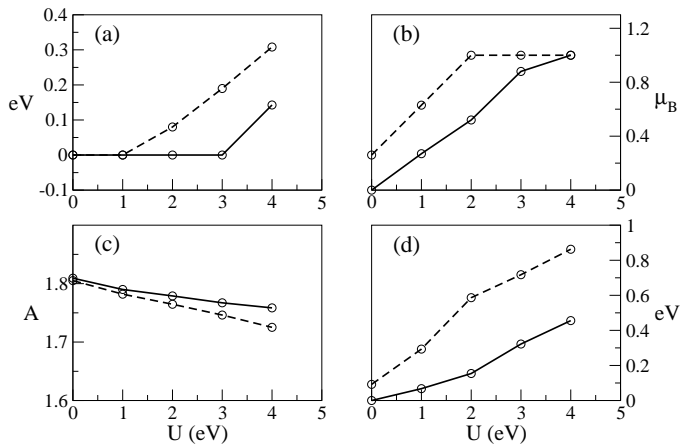}
\end{center}
\vspace{-0.5cm}
\caption[]{Sn/Si(111) (dashed) and Sn/Ge(111) (solid) $\sqrt3$ calculations 
for increasing correlation parameter $U$: (a) minimum band gap, (b) spin moment 
per adatom, (c) Sn vertical height measured over the outer semiconductor plane, 
(d) magnetic exchange splitting. Lines are guides for the eye.}
\label{fig1}
\end{figure} 

Within the last few months that situation changed drastically. Cooling 
Sn/Ge(111) below 20 K apparently turns it from a $3\times3$ distorted metal to an 
undistorted $\sqrt3$ insulator, presumably a MI \cite{cortes}. 
Equally striking, Sn/Si(111) is now shown by Modesti's group \cite{modesti}   
to turn continuously from an undistorted metal to a narrow gap insulator, 
again presumably a MI, below 60 K. If Sn/Ge(111) is intriguing enough 
due to the disappearance of structural distortion at low T, the result 
on Sn/Si(111) is no less puzzling. Low temperature 
photoemission \cite{uhrberg, modesti} 
finds Sn-related surface bands below $E_F$ at the $\Gamma$-point, as 
though folded over from the $K$-point. This is an intriguing but clear indication 
of $3\times3$  periodicity, for in the $\sqrt3$ surface there are no such filled
surface states at $\Gamma$ \cite{profeta}. Structural tools including 
scanning tunneling microscopy (STM) and photeoelectron diffraction show 
only $\sqrt3$ periodicity, so that the $3\times3$ motif is not structural. For 
both surfaces therefore, the LDA predictions of metallic ground states turn 
out to be in error.

Unraveling this situation calls for a renewed theoretical effort. Calculations
should at the same time be of first principles quality, so as to permit 
total energy comparisons, but also treat correlations 
more accurately than LDA, so as to identify MIs if and when present. The LDA+U 
approach, while still a mean field approximation (thus for example replacing a MI 
with a fictitious magnetic band insulator) does satisfy these 
criteria \cite{anisimov} and is suitable to describe quantitatively
surface MIs \cite{anisimov1}. We therefore conducted a series of accurate LDA 
and LDA+U calculations for Sn/Si(111), and for Sn/Ge(111).  

\begin{figure}
\begin{center}
\includegraphics[width=0.45\textwidth,clip]{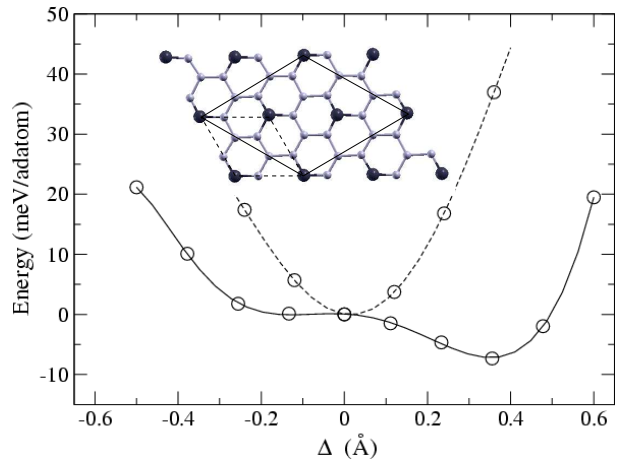}
\end{center}
\vspace{-0.5cm}
\caption[]{Sn/Ge(111): Total energy per adatom as a function of $3\times3$
up-down distortion $\Delta$. LDA (solid line) and LSDA+U, $U$=4 eV (dashed line).
The zero of energy is at $\Delta=0.0$ \AA\ in both cases.
Inset: Schematic of $\sqrt3\times\sqrt3R30^{\circ}$ unit cell (dashed line)
and of $3\times3$ unit cell (solid line). Black spheres represent Sn adatoms,
grey spheres Si (Ge) atoms of the first and second layer.}
\label{fig2}
\end{figure}

The geometries considered were periodic slabs consisting of three (111) semiconductor 
bilayers, H-saturated at the bottom, and with the 1/3 monolayer of Sn adsorbed at
$T4$ sites on the top surface. We used the plane wave implementation of 
density functional theory \cite{espresso, ballabio} in the gradient corrected local spin density approximation 
(LSDA), extended to include a Hubbard U (LSDA+U). Beginning with an undistorted $\sqrt3$ 
geometry, (one Sn per cell) we turned on an increasing electron-electron
onsite repulsion $U$ ranging from zero to an estimated full value of $U\simeq4$ eV\cite{note_U} 
for two electrons occupying  the same Sn $5p_z$ orbital -- this single orbital 
constituting about 50\% of the surface state \cite{degironc}.
The effect of a finite $U$ parameter in LSDA+U is to favor integer occupancy of this 
orbital, which is the crucial effect of strong correlations.
 
Fig.\ref{fig1} summarizes the main physical quantities calculated at zero-temperature 
for  Sn/Si(111) and Sn/Ge(111), illustratively shown for increasing $U$ values. 
Sn/Si(111), metallic and weakly magnetic already at $U\simeq0$, 
develops a stronger 
magnetization and eventually a metal insulator transition $U\simeq2$ eV, where the 
spin moment per adatom site saturates to $1\mu_B$. The insulating gap reaches 0.3 eV 
at the final realistic value $U\simeq4$ eV, where the exchange splitting between 
opposite spins orientations becomes as large as 0.8 eV. The insulating state gains 
Coulomb energy, but loses band energy relative to the metallic state. This is also 
reflected indirectly by the predicted downward geometrical relaxation of the Sn 
adatoms by about 0.06 \AA\ closer to the Si substrate, upon going from metal at to insulator. 
Vertical adatom positions control hybridization between the $5p_z$ dangling bond 
orbital and the underlying Si-Si antibonding state \cite{degironc, ballabio1}. That 
hybridization is strong in the metallic state (and actually modulated in the 
$3\times3$ distortion), but counterproductive and thus weaker in the insulating state. 
Within the artificial $\sqrt3$ geometry, results for Sn/Ge(111) are on the whole similar 
to those in Sn/Si(111), except that magnetism does not develop at $U\simeq0$, and 
the metal-insulator transition only occurs for $U\simeq4$ eV. All gaps and splittings 
are correspondingly smaller.

We may crudely identify the magnetic 2D insulators near $U\simeq4$ eV 
with the actual MI states of the true surfaces, whose measured gaps are 
qualitatively similar, ranging from zero to hundreds of meV \cite{cortes, modesti}. 
Before taking that identification 
seriously, we must however compare the structural properties of the 
insulator with those of the competing phases, 
in particular with the $3\times3$ distorted metallic ones.
To that end we repeated all LSDA+U calculations in an enlarged $3\times3$ cell, where
both distortive and ferromagnetic order parameters are allowed. The outcome
for Sn/Si(111) was uneventful, and identically the same undistorted magnetic states
were recovered for all $U$ values. That confirms that a $\sqrt3$ MI is indeed the  
LSDA+U predicted ground state of Sn/Si(111). Apart from magnetism, not yet investigated
experimentally, this fully agrees with recent data by Modesti {\em et al.}
 \cite{modesti}. 

The evolution is different in Sn/Ge(111) where initially at $U\simeq0$
the ground state is metallic, nonmagnetic, but now $3\times3$ distorted, one
Sn up by $\Delta /2$, two down by $\Delta /2$. The up-down amplitude 
$\Delta\simeq0.36$ \AA\ and the energy gain $\sim$ 9 meV/Sn are same as  
previously found long ago by similar methods 
\cite{avila, perez, degironc, ballabio} 
and hitherto believed to describe the true ground state of this surface.
However as shown in Fig.\ref{fig2} (obtained by constrained structural 
optimization along the distortion path) for increasing $U$ the distortion $\Delta$ 
eventually disappears giving way to the same magnetic insulator previously 
found at $U\simeq4$ eV. Hence inclusion of correlations strongly 
modifies the ground state of Sn/Ge(111) from a $3\times3$ distorted metal
to a magnetic insulator with $\sqrt3$ structural symmetry. 
This agrees with the STM and photoemission results 
of Cortes {\em et al.} below 20 K \cite{cortes}.
 
\begin{figure}
\begin{center}
\includegraphics[width=0.45\textwidth,clip]{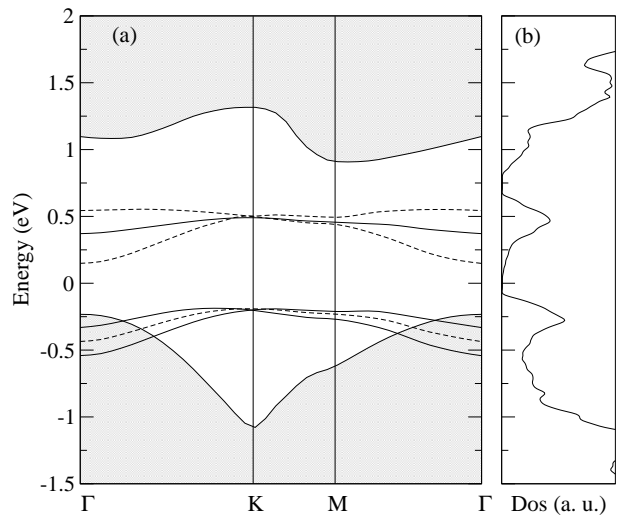}
\end{center}
\vspace{-0.5cm}
\caption[]{(a) Sn/Si(111): LDA+U spin-up bands (solid
line) and spin-down bands (dashed line) for the ferrimagnetic 
$3\times3$ phase. (b) Total density of electronic states (solid line)
 showing the gap at the Fermi level (zero).}
\label{fig3}
\end{figure}

In the MI phase, an unpaired electron is localized near each
Sn adatom site. Ignoring anisotropy, these 1/2 spins form a 2D triangular
Heisenberg-like magnet with intersite antiferromagnetic  (AF) exchange coupling whose order
of magnitude is set by $J \sim t^2/U_{eff} \sim5$ meV (for $t \sim0.05$ eV, $U_{eff} 
\sim0.5$ eV \cite{scandolo}). Their ground state could thus be either a $120^{\circ}$  
noncollinear Neel state \cite{huse} or a spin liquid \cite{morita}. 
We made use of collinear LSDA+U  to calculate the actual $J$ values for
Sn/Si(111) and Sn/Ge(111). We carried out two separate $3\times3$ unit cell $U=4$ eV 
calculations, one fully ferromagnetic (3$\mu_B$ per cell), the other {\it ferri}magnetic 
(1$\mu_B$ per cell), with two adatom spins up, one down \cite{footnote1}. 
The total energy change per $3\times3$ cell $\Delta E$
amounts to switching 6 bonds from antiferromagnetic to ferromagnetic,
which for spins 1/2 implies an energy loss of 9$J$. 
We calculated $\Delta E$=27 and 42 meV, 
implying $J \sim$ 3 and 5 meV for Sn/Si(111) and Sn/Ge(111) respectively.
Because of the better accuracy of the ferro calculation, these
are most likely underestimates, and we conclude that at least up to a hundred 
Kelvin the MI state of Sn/Si(111) and Sn/Ge(111) is likely to possess 
AF short range order, probably even outright Neel order, 
although renormalized by quantum fluctuations \cite{huse}. 
Hole spectral function calculations in the 2D 
triangular t-J model \cite{manuel} suggest that purely 
magnetic $3\times3$  order will give rise to photoemission features 
typical of a $3\times3$ band folding, even in the absence of
structural $3\times3$ periodicities. Fig.\ref{fig3}  shows the LSDA+U 
electronic bands of Sn/Si(111) ferrimagnetic state which exemplify that folding.  
The photoemission intensity near $E_F$ at $\Gamma$ \cite{uhrberg, 
modesti} observed in Sn/Si(111) is thus likely due to $3\times3$ AF order 
and consequent folding. The same $3\times3$ magnetic folding should now be
searched and detected at the true 
$\Gamma$-point of Sn/Ge(111)'s  low temperature MI state. 


These results are intriguing, and lead on to a number of interesting questions. 

First, why do the MI states evolve into metallic phases,
beginning near 60 K and 20 K respectively in Sn/Si(111) and in Sn/Ge(111)?
Moreover, why do these finite temperature metallic phases resemble 
the uncorrelated LDA ground states calculated at T=0?
%

\begin{figure}
\begin{center}
\includegraphics[angle=-0,width=0.4\textwidth,clip]{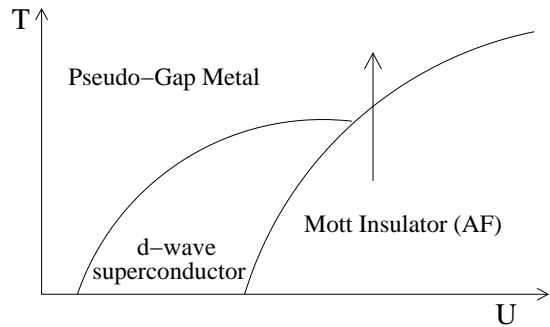}
\end{center}
\vspace{0cm}
\caption[]{Schematic phase diagram of Sn/Si(111) in the temperature-interaction 
plane. The arrow describes the entropy-induced metallization due to 
AF order in the MI.}
\label{fig4}
\end{figure}

The driving force for the observed T-induced metallization of MI states is,  
in analogy with $V_2O_3$ and 2D $k-(ET)_2Cu[N(CN)_2]Cl$ \cite{kanoda} 
their lack of spin entropy, frozen out by AF short range order. 
If $\Delta E \sim W(U/U_c-1)$ represents a typical insulator--metal energy 
difference per site (for a transition at $U = U_c$), and 
$\Delta S(T) \sim \gamma T^2$ the entropy-related free energy difference
(assuming AF frozen spins and a charge gap in the MI), 
then the low--T MI--metal 
phase boundary is predicted of the form:
\begin{equation}
T_{im} = [W (\frac{U}{U_c} -1)/\gamma]^{\frac{1}{2}}
\end{equation}
which is sketched in Fig.\ref{fig4}, and well brought out 
experimentally in 3D compounds.\cite{kanoda}
Here $W \sim 0.3$ eV is the bare bandwidth, and $\gamma$ 
is the electronic specific heat coefficient 
(perhaps of order 0.01 J/mol $K^2$, or roughly 0.1 meV/site $K^2$ 
as in 2D organics\cite{kanoda}). 
We thus suggest that quasiparticle entropy 
drives these surfaces across the insulator -- metal 
phase boundary, resurrecting the metallic phase, itself only metastable 
at $T$ = 0. Four point conductance measurements\cite{wells} could be  
of great help in ascertaining this scenario.
  
The next and crucial if still speculative question is whether there 
is any chance to realize a 2D
(power-law ordered) superconducting state in these surfaces.\cite{bertel} 
As indicated in Fig.\ref{fig4}, in the general
phase diagram of the triangular Hubbard lattice, as realized 
by organics under pressure\cite{kanoda} -- 
there is indeed a low--T $d$-wave superconductor on the metallic side next 
to the MI phase. We thus propose that one should 
try to achieve superconductivity by metallization of Sn/Si(111) below 60 K.
Metallization could be attempted by $e.g.$, reducing $U$, or by increasing 
the bandwidth, or by doping surface bands away from half-filling. 
The latter might be realized by alkali deposition; the former possibly 
through heavy doping of the Si bulk  substrate. We note in passing that 
the $d$-wave superconducting state in a triangular 
lattice would probably break rotational symmetry \cite{yunoki}, an event 
readily observable by STM.

One remaining unknown is the role of spin orbit coupling. While large 
in Sn, the largely $p_z$ nature of the state will reduce its relevance. 
Possibly, some amount of magnetocrystalline anisotropy will result. 
Depending on its  sign, the spin 1/2 sites will turn from Heisenberg to 
either Ising (out of the surface plane; in this case, the ferrimagnetic state 
considered above actually corresponds to the true MI ground state) or XY 
(in the surface 
plane). This aspect remains open for future investigation.

In conclusion, in this Letter we provide a first theoretical background for the
insulating ground state just observed in surface $\alpha$-phases Sn/Si(111)
and Sn/Ge(111), and obtain quantitative indications that the insulating phase
observed in both should be of correlation origin. LSDA+U calculations predict a stable
magnetic and (Mott-Hubbard like) insulating $\sqrt3$ phase in Sn/Si(111), 
as well as the disappearance of the $3\times3$ distortion in Sn/Ge(111), 
 as is experimentally observed. Temperature induced metallization is argued 
to represent evidence for antiferromagnetism in the Mott state. 
Should metallization be provoked at sufficiently low temperatures, a 
2D $d$-wave superconducting state could be achieved. 
A number of experimental approaches, including surface doping, 
and four point conductance measurements are suggested for the future.

We acknowledge rich exchanges of ideas and information with S. Modesti, and 
illuminating discussions with M. Fabrizio, G.E. Santoro, S. Sorella, and 
G. Baskaran. This work was sponsored 
by MIUR FIRB RBAU017S8 R004, FIRB RBAU01LX5H, MIUR COFIN 2003, PRIN/COFIN2004, and
PRIN/COFIN2006.

\end{document}